\documentclass[superscriptaddress,amsmath,amssymb,aps,prl,twocolumn]{revtex4-2}

\usepackage{graphicx}
\usepackage[colorlinks]{hyperref}
\usepackage{caption}
\usepackage{subcaption}
\usepackage{float}
\usepackage[english]{babel}
\usepackage{placeins}

\begin{document}

\title{Quantum-Inspired Multispectral Range Finding}

\author{Autumn Rose Landwehr\textsuperscript{*}}
\affiliation{Department of Physics and Astrophysics, University of North Dakota, Grand Forks, North Dakota, USA}
\email{autumn.landwehr@und.edu} 
\author{Thilini Bamunu Arachchige}
\affiliation{Department of Physics and Astrophysics, University of North Dakota, Grand Forks, North Dakota, USA}
\author{Elaganuru Bashaiah}
\affiliation{Department of Physics and Astrophysics, University of North Dakota, Grand Forks, North Dakota, USA}
\author{Markus Allgaier}
\affiliation{Department of Physics and Astrophysics, University of North Dakota, Grand Forks, North Dakota, USA}

\begin{abstract} 
Optical remote sensing instruments either resolve the world in three dimensions through laser altimetry, or in two spatial dimensions and spectrum in hyperspectral imaging, but never in all four degrees of freedom simultaneously. Vertically resolved hyperspectral imaging could resolve complex phenomena in the ocean, atmosphere, ecosystems, and cryosphere. We present a multispectral LiDAR scheme that uses dispersion-assisted time-multiplexed spectroscopy commonly used in quantum optics. We show a tabletop implementation with different spectral separations and range windows with independent ranging of spectral components. The scheme can in principle work at ultra-long range and large spectral bandwidth.
\end{abstract}

\maketitle

\section{Introduction}
Spaceborne optical remote sensing instruments measure our environment in several of four dimensions: Three spatial, and one spectral. Typical imaging instrument resolve two spatial and the spectral dimensions, while laser altimeters resolve all three spatial dimensions, but no current spaceborne instrument can comprehensively resolve all 4 degrees of freedom. Such an instrument - a multispectral laser ranging or LiDAR (light detection and ranging) instrument - would not only provide the functionalities of a laser altimetry and hyperspectral imager in the same instrument for reduced size, weight, and power requirements (SWaP), but provide vertically resolved hyperspectral information that could enable novel insights in atmospheric, ocean, biosphere, and cryospheric science. One example for a system in nature that is frequently interrogated with both laser altimetry and hyperspectral imaging is vegetation \cite{VERRELST2015273,DUBAYAH2020100002}. Laser ranging at two wavelengths is already in use for greenhouse gas mapping with differential absorption LiDAR (DIAL) \cite{Romanovskii2025}.
Here, we propose that such an instrument could be based on photon-counting time-of-flight LiDAR \cite{gariepy2015single,shrestha2009performance,liu2023compact}, just like NASA's Advanced Topographic Laser Altimeter System (ATLAS) onboard ICESat-2 \cite{markus2017ice,NEUMANN2019111325,NEUENSCHWANDER2019247}. While frequency-modulated and amplitude modulated continuous wave LiDAR uses interferometric detection at the receiver \cite{royo2019overview,bianconi2025requirements}, time-of-flight LiDAR incoherently detect backscatter and derives range from the round trip time, which works at the single photon level. If the laser repetition rate is low or the range window is much shorter than the maximum range (like it is the case for ATLAS), we propose to use the time between laser pulses to measure the spectral domain. Quantum optics experiments resolve the spectrum at the single photon level with frequency to time mapping. At the receiver, light passes through a dispersive optical fiber or fiber Bragg grating that introduces a large amount of group delay \cite{avenhaus2009fiber,Davis:17}, causing different wavelengths to arrive at different times. This allows us to detect all spectral components with a single detector instead of using one detector for each spectral band \cite{qian2023design}. Similar concepts have been explored \cite{Ren:18,Jiang2020TimeStretchLiDAR,Chens24175741}, but the possibility of independently ranging all spectral components and the implications on the achievable range and range window have not been explored.

\begin{figure}
    \includegraphics[width=0.91\columnwidth]{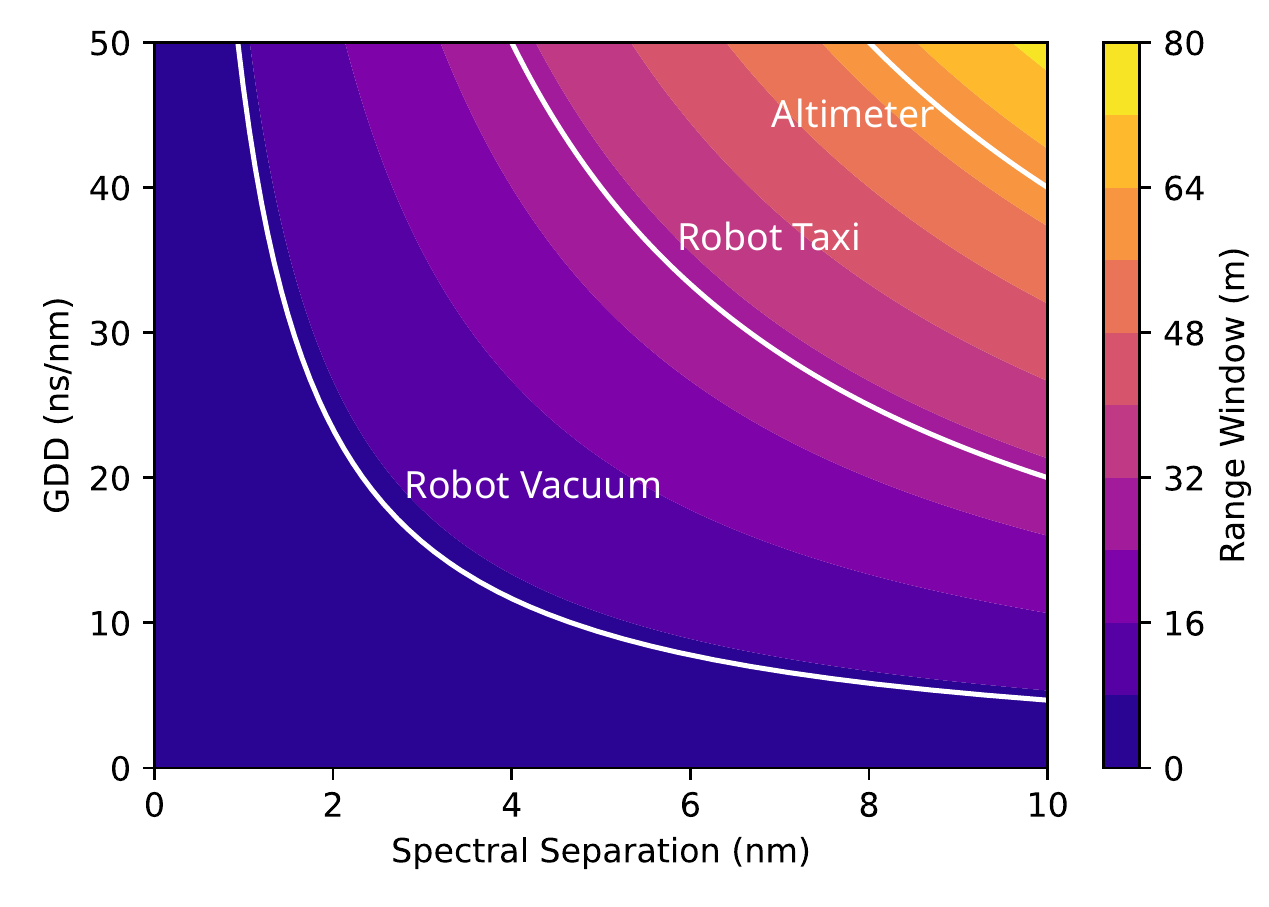}
    \caption{Range window as a function of GDD and spectral separation. The solid white lines show approximate range windows for common time-of-flight LiDAR applications: 6 m for consumer robot vacuum, 30 m for self-driving vehicle, 60 m for laser altimeter.}
    \label{fig:heatmap}
\end{figure}

In this letter, we describe the concept of a quantum-inspired multispectral LiDAR with frequency-to-time mapping and show a tabletop demonstration that uses a mode-locked broadband laser source, a programmable filter, and commercially available dispersive fiber modules to supply group delay dispersion. We achieve a maximum range of 7.57 m, with an independent range window for each spectral component of the order of 1 m. We describe the tradeoffs between maximum range, range window, and spectral resolution.

\begin{figure*}
    \centering
    \includegraphics[width=0.95\textwidth]{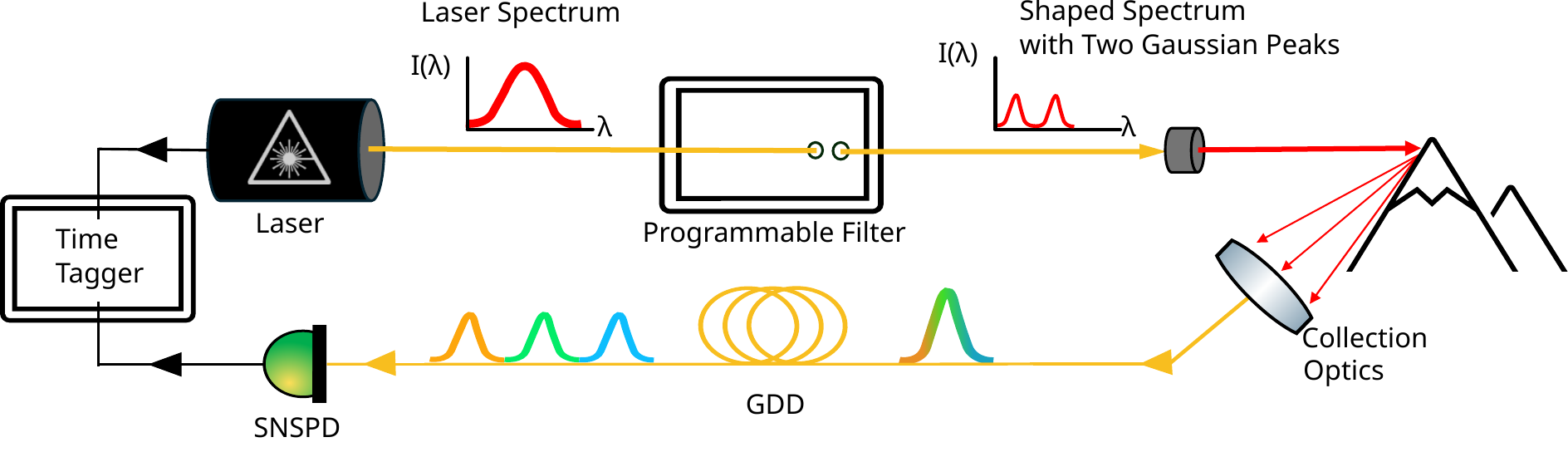}
    \caption{Schematic for the experimental demonstration of multispectral LiDAR utlilizing a dispersion-compensating fibers to resolve spectrum in time. The mode-locked laser is free-running and pulsing at a rate of 100 MHz, which is registered by the Time Tagger. The laser spectrum is shaped into two well-separated gaussian spectra. Emitted light is directed through a borehole through the off-axis parabolic collection mirror, so that backscatter is collected on the direct backwards path. The wavepacket containing two gaussian peaks in spectrum is directed through the dispersion compensation modules, which separates it into two gaussian wavepackets in time with different central frequency before being registered on the superconducting nanowire single photon detector (SNSPD).}
    \label{fig:schematic}
\end{figure*}

\section{Dispersion-Assisted Multispectral Measurements}

Long range LiDAR instrument such as the photon-counting Advanced Topographic Laser Altimeter System onboard ICESat-2 \cite{NEUMANN2019111325} operate at a range of the order of 500 km while only opening a range window of the order of 100 m around earth's surface, leaving almost 100 \textmu s of ''unused`` time between laser pulses. Many quantum optics experiments involving photon pairs entangled in time and frequency make use of a similar time gap between laser pulses to implement spectroscopy at the single photon level by mapping relative frequency to the time domain using with dispersion in optical fibers \cite{avenhaus2009fiber} or fiber Bragg gratings \cite{Davis:17}. There are two fundamentally different use cases for multispectral LiDAR with different constraints on the range window and spectral separation:
\begin{itemize}
    \item Single return per pulse from a hard target: If all backscatter originates from the same opaque target, then all spectral components have the same timing at the target. The spectral components only need to be separated enough to be resolvable at the receiver, similar to experiments in quantum optics. Spectral resolution is a function of GDD and and temporal resolution. The measurement resolves the reflectance spectrum of a single target. Full range is accessible to range window.
    \item Multiple returns with independent reflectance/transmittance windows: Here, any combination of spectral components could scatter back from anywhere in the range window, so they need to be independently ranges. Spectral components need to be separated in time by at least as much as the anticipated range window. The measurement resolves a range-dependent reflectance spectrum.
\end{itemize}
We treat the second, more challenging case here in more detail. Photons experience group delay delay $GDD$ at the receiver, introducing a time delay $\Delta t$ between spectral lines separated by $\Delta\lambda$. If used in LiDAR where we wish to independently range two spectral components, that time delay is equivalent to the range window:
 
\begin{equation}
    \Delta t = \dfrac{GDD \Delta\lambda c }{2}.
\end{equation}

Here, c is the speed of light in air. Figure \ref{fig:heatmap} shows the range window as a function of GDD and the separation between spectral bands, showing that reasonable range windows of the order of  tens of meters are theoretically achievable. It should be noted that if we relax the assumption of independent ranging and instead assume that all backscatter originates from the same hard target, we would only require a sufficiently large temporal separation to resolve the spectral components, and the range window remains unchanged from an equivalent single wavelength system. As long dispersive fibers add considerable attenuation, the range of such a LiDAR system would be reduced by the factor:

\begin{equation}
    R^- = \sqrt{T} = \sqrt{\frac{{ I(r_{red)}}}{I(r_0)}}  = \sqrt{10^{\frac{-A(l)}{10}}} \ ,
\end{equation}

\noindent where $r_0$ is the range without attenuation, $r_{red}$ is the new reduced range. Where $A(l)$ is the attenuation in the system in dB as a function of fiber length $l$, and $T$ is the transmittance.

\section{Experimental Implementation}

A tabletop proof-of-principle demonstration of this technique was implemented, as outlined in Figure \ref{fig:schematic}. A pulsed laser source (Vescent \textit{RUBRIComb}) with a repetition rate of 100 MHz and a broadband spectrum centered at 1550 nm is used. The desired spectrum is obtained using a programmable filter (\textit{Coherent Waveshaper 1000A}), in which the applied attenuation along the spectrum is programmed to achieve a spectrum consisting of two gaussian peaks. Adjustable parameters include the bandwidth and spectral separation. The light is then directed towards the target at known range, and backscattered light is collected into a single-mode optical fiber with an off-axis parabolic mirror. After the signal is coupled into the fiber, dispersion compensation modules (DCMs) are used to resolve the signal by providing group delay dispersion (GDD), which creates a difference in group delay for the two gaussians centered at different wavelengths. The DMCs used here are based on dispersion compensating fibers. The experiment contains four DCMs, two providing 0.36 ns/nm GDD and two providing 1.44 ns/nm GDD. Any combination of the DCMs can be used to provide GDD from 0.36 to 3.6 ns/nm. The signal is detected using \textit{Quantum Opus One} superconducting nanowire single photon detectors (SNSPDs). Time of pulse emission and relative time of arrival of photons is measured using \textit{Swabian Instruments Time Tagger Ultra}. Relative time of arrivel relative to actual range is calibrated with a target at known range. 

\begin{figure}
    \includegraphics[width=0.95\columnwidth]{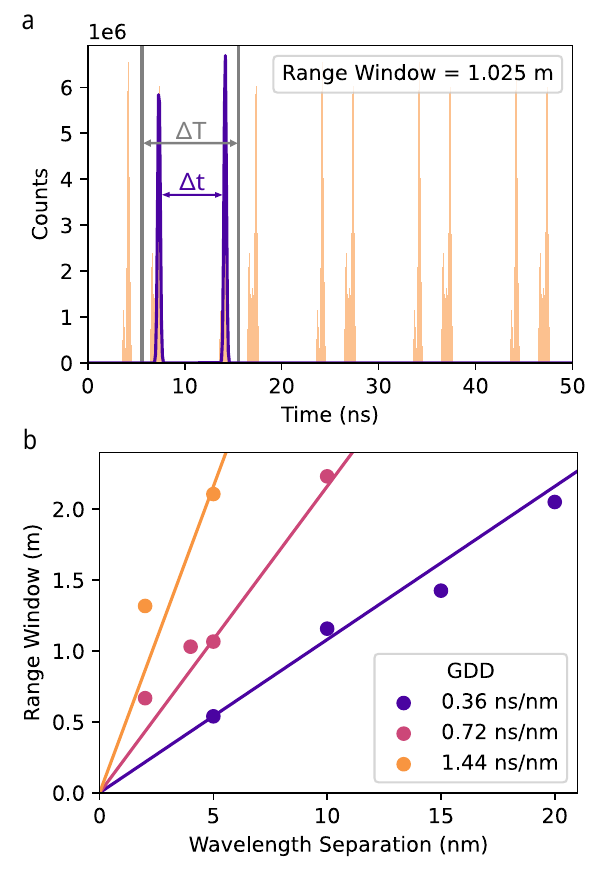}
    \caption{(a) Sample spectrum where 0.36 ns/nm dispersion is used to resolve two spectral lines of separation $\Delta\lambda$ = 20 nm. Duration of 50 ns, or five laser pulses. The solid orange line is the spectrum, the solid purple line is the gaussian fit. Solid gray lines denote the ambiguity range of 10 ns. (b) Measured range window values as a function of spectral separation for three different GDD values. Solid lines correspond to expected range window.}
    \label{fig:rangewindow}
\end{figure}

A sample spectrum of two gaussian-shaped peaks resolved in the time domain is shown in Figure \ref{fig:rangewindow}. Here, the spectral separation is $\Delta\lambda$ = 20 nm, and the spectrum was resolved using 0.36 ns/nm GDD. First, the system's maximum range $\Delta$$T$ is determined using a gaussian located at the center frequency of the laser (1550nm): $\Delta$$T$ is the separation of adjacent pulses centered at the value of the mean fit for a gaussian at this center frequency, which is conducive with the ambiguity range dictated by the laser system's repetition rate. Next, the maximum possible range window $\Delta$$t$ for a set of $\Delta\lambda$ and GDD is determined by the difference of the mean values of gaussian fits to the spectral lines, which is converted to a distance scale. Note that if the range window exceeds the maximum range, adjacent signals will begin to overlap. For the case of independent ranging of the two wavelengths, it is not assumed that both returns originate from the same target. For same target, the range window is unchanged by using multiple wavelengths.

\begin{figure}
    \includegraphics[width=0.95\columnwidth]{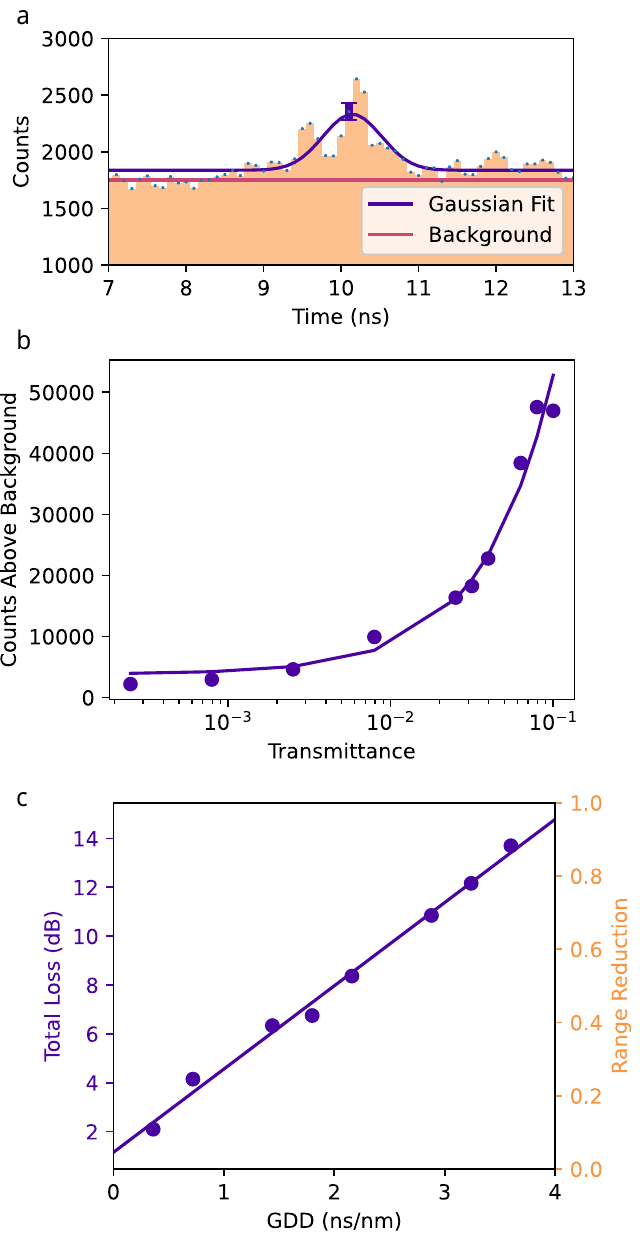}
    \caption{(a) Amplitude error bar of gaussian fit and noise floor are used to calculate SNR. The magenta line is fitted to the noise floor where no return light is present, the solid purple line is the gaussian fit to the return, measured at 60 s integration time (b) Signal-to-noise ratio (SNR) as a function of transmittance (c) Total loss and range reduction as a function of GDD.}
    \label{fig:uncertainty}
\end{figure}

The intensity of the return signal is affected by the ranging distance as well as the amount loss introduced by the DCMs, as seen in Figure \ref{fig:uncertainty}. The signal-to-noise ratio (SNR) determines the range limit and range reduction. SNR was defined as the uncertainty of the amplitude fit parameter of the gaussian to noise floor.  To model the effect of increasing the range, neutral density (ND) filters were used.

\begin{equation}
T = 10^{-OD} \propto \frac{1}{r^2}
\end{equation}

where $OD$ is the optical density, $T$ is transmittance, and $r$ is the range. Loss is also introduced with increasing GDD of the DCMs, as described in equation 2. The equivalent range at the largest introduced loss is 7.57 m. In Figure \ref{fig:uncertainty}c we show the system loss introduced by adding the DCMs as a function of provided GDD, and the corresponding reduction in range. For a GDD of 4 ns/nm, range is effectively reduced by one order of magnitude, showing that independent ranging of spectral components, which requires tremendous amounts of GDD to achieve even modest range windows, would likely not be feasible for spaceborne sensors.

\section{Conclusion}

We show a scheme and proof-of-principle experiment on multispectral LiDAR that achieves spectral resolution with simultaneous ranging at the single photon level through dispersion-assisted time of flight spectroscopy more commonly used in quantum optics. The scheme allows for independent ranging of the wavelength and can in principle support long distance ranging and hyperspectral resolution over a large spectral bandwidth for use in future remote sensing instruments. While our tabletop demonstration is limited by the laser system's fast repetition rate, which introduces a short ambiguity range of just 1.5 meters, lower repetition rate systems that also use SNSPDs have been demonstrated with ranges of of 200 km \cite{Li:21} and even 1000 km \cite{https://doi.org/10.1029/2024SW003977}, both using setups similar in nature to the one used here, but lacking the dispersive spectral resolution.

\section*{Funding}

This project was funded by the University of North Dakota, and the National Science Foundation under Grant No. 2427047.

\section*{Acknowledgment}

We thank Cody C. Payne for fruitful discussions on pulse shaping.

\section*{Disclosures}

The authors declare no conflicts of interest.

\section*{Data availability}

All produced data is part of the manuscript. The code and raw data are available from the corresponding author upon reasonable request.

\bibliography{References}

\end{document}